\documentclass[preprint, preprintnumbers, amsmath ,amssymb, letterpaper, superscriptaddress]{revtex4}

\usepackage{amsmath}

\usepackage{graphicx}
\usepackage{dcolumn}
\usepackage{bm}

\usepackage{subfigure}

\preprint{}

\begin{document}

\title{\bf Envelope Hamiltonian for Charged-Particle Dynamics in General Linear Coupled Systems}

\author{Moses Chung} \email{mchung@unist.ac.kr}
\affiliation{Department of Physics, Ulsan National Institute of Science and Technology, Ulsan 689-798, Korea}
\author{Hong Qin}
\affiliation{Plasma Physics Laboratory, Princeton University, Princeton, New Jersey 08543}
\affiliation{Department of Modern Physics, University of Science and Technology of China, Hefei, Anhui 230026, China}
\author{Ronald C. Davidson}
\affiliation{Plasma Physics Laboratory, Princeton University, Princeton, New Jersey 08543}

\date{\today}


\begin{abstract}
We report the discovery of an
envelope Hamiltonian describing the charged-particle dynamics in general linear coupled lattices.
\end{abstract}

\maketitle


The most fundamental theoretical tool in designing and analyzing
an uncoupled lattice system is the well-known Courant-Snyder (CS) theory \cite{CS}.
Almost all beam and accelerator physics textbooks begin a discussion of the charged particle beam dynamics in terms of the CS theory.
The main components of
the CS theory are the envelope equation, the phase advance, the transfer matrix, and the CS invariant.
While formulated on the basis of the single-particle equation of motion,
these physical quantities provide an effective and elegant means to describe the motions of the collection of charged particles making up the beam.

For example, for a given lattice with focusing coefficient $\kappa_q (s)$ in the $x$-direction,
the single-particle dynamics are governed by the oscillator equation \cite{Davidson.book1}
\begin{equation}
x'' + \kappa_q (s) x= 0,
\label{1D_oscillation}
\end{equation}
where $x(s)$ is the transverse displacement of a beam particle about the reference orbit, and $s$ is a scaled time variable with dimensions of length.
Transforming Eq. (\ref{1D_oscillation}) according to $x(s) = A_x w \cos \left[ \phi (s) + \phi_0 \right] $,
where $A_x$ and $\phi_0$ are constants, and the phase advance $\phi = \int_0^s ds' / w^2(s')$ \cite{Davidson.book1},
the corresponding envelope function $w(s)$ evolves according to
\begin{equation}
w'' + \kappa_q (s) w  = w^{-3}.
\label{1D_envelope}
\end{equation}
For a given beam emittance,
$w(s)$ provides the information on the transverse excursion amplitude of the beam particle in configuration space.
We note that there is an additional nonlinear term $w^{-3}$ in the envelope equation (\ref{1D_envelope}),
which prevents a change in the sign of $w(s)$ \cite{Accelerator_physics2}.

The solution of Eq. (\ref{1D_oscillation}) can be expressed as a symplectic linear map
that advances the phase space coordinates
\begin{equation}
\left(
  \begin{array}{c}
    x   \\
    x' \\
  \end{array}
\right)
=
\left(
         \begin{array}{cc}
           w & 0 \\
           w' & w^{-1} \\
         \end{array}
       \right)
       P^{-1}
       \left(
         \begin{array}{cc}
           w^{-1} & 0 \\
           -w'    & w \\
         \end{array}
       \right)_0
\left(
  \begin{array}{c}
    x\\
    x' \\
  \end{array}
\right)_0,
\label{M(s)_1D}
\end{equation}
where subscript ``0" denotes initial conditions at $s = 0$ and $P$ is
the phase advance matrix, which is determined by the following differential equation with the initial condition $P_0 = I$:
\begin{equation}
P ' = P \left(
          \begin{array}{cc}
            0            & - w^{-2} \\
            w^{-2} & 0 \\
          \end{array}
        \right).
        \label{P_1D}
\end{equation}
Here, $w^{-2}$ is the phase advance rate.
In the original CS theory, the solution for $P$ is trivial, and it is given by the rotation matrix
\begin{equation}
P =
\left(
      \begin{array}{cc}
        \cos \phi & - \sin \phi \\
        \sin \phi & \cos \phi \\
      \end{array}
    \right).
\end{equation}

The deeper connection between the single-particle equation of motion (\ref{1D_oscillation}) and
the envelope equation (\ref{1D_envelope}) can be investigated using the  Hamiltonian formulation.
The Hamiltonian corresponding to  Eq. (\ref{1D_oscillation}) is given by
\begin{equation}
H = \frac{1}{2} p_x^2 + \frac{1}{2} \kappa_q(s) x^2,
\end{equation}
where $p_x$ is the scaled momentum.
Often the Hamiltonian is conveniently expressed in the matrix form
\begin{equation}
H = \frac{1}{2}
\left(
  \begin{array}{cc}
    x, & p_x \\
  \end{array}
\right)
                        \left(
                          \begin{array}{cc}
                            \kappa_q & 0 \\
                            0   & 1 \\
                          \end{array}
                        \right)
\left(
  \begin{array}{c}
    x   \\
    p_x \\
  \end{array}
\right).
\end{equation}
Then, the equations-of-motion are given by
\begin{eqnarray}
x'   &=&  \frac{\partial H }{\partial p_x } = p_x,  \label{EM1} \\
p_x' &=& -  \frac{\partial H }{\partial x } = - \kappa_q(s) x. \label{EM2}
\end{eqnarray}
The corresponding envelope functions are determined from
\begin{eqnarray}
w' &=& v, \label{ENV1} \\
v' &=& - \kappa_q(s) w + w^{-3} \label{ENV2}.
\end{eqnarray}
Different from the original CS theory, we have expressed the envelope equation (\ref{1D_envelope}) in terms of the two first-order differential equations
in Eqs. (\ref{ENV1}) and (\ref{ENV2})
in order to indicate that the envelope function $w$ and its corresponding momentum $v$ form
a certain Hamiltonian structure \cite{Accelerator_physics3}.
Indeed, we  immediately note that there exits an envelope Hamiltonian
\begin{eqnarray}
H_{env}
&=& \frac{1}{2} v^2 + \frac{1}{2} \kappa_q(s) w^2 + \frac{1}{2} w^{-2} \nonumber \\
&=& \frac{1}{2}
\left(
  \begin{array}{cc}
    w, & v \\
  \end{array}
\right)
                        \left(
                          \begin{array}{cc}
                            \kappa_q & 0 \\
                            0   & 1 \\
                          \end{array}
                        \right)
\left(
  \begin{array}{c}
    w   \\
    v \\
  \end{array}
\right)
+ \frac{1}{2} w^{-2},
\label{Henv1}
\end{eqnarray}
which yields the envelope equations (\ref{ENV1}) and (\ref{ENV2})
through the Hamiltonian formulation
\begin{eqnarray}
w'   &=& \frac{\partial H_{env} }{\partial v }, \\
v'   &=& - \frac{\partial H_{env} }{\partial w }.
\end{eqnarray}
Further, we introduce the effective envelope potential $V_{env}$ defined as
\begin{equation}
V_{env} = \frac{1}{2} \kappa_q(s) w^2 + \frac{1}{2} w^{-2}.
\end{equation}
The existence of the envelope Hamiltonian and potential provides
the idea that, in certain circumstances, beam matching or optimization of beam transport could be
achieved by finding the equilibrium solution of the envelope Hamiltonian
(see, for example, Ref. \cite{Accelerator_physics3}).


Attempts to extend the original CS theory to the cases of general linear coupled lattices have a long history.
Nonetheless, no single method has yet been adopted as a {\it de facto} standard in the beam physics community.
The recently developed generalized CS theory \cite{GCS_PRL, GCS.PRSTAB} for the single-particle dynamics
is particularly noteworthy in the sense that
it retains all of the elegant mathematical structures of the original CS theory
with remarkably similar physical meanings.
The envelope function is generalized into an envelope matrix (i.e., $w$ is now a $2\times 2$ matrix),
and the phase advance is generalized into a 4D symplectic rotation.
Furthermore, the generalized theory includes not only all of the linear elements (i.e., quadrupole, skew-quadrupole, and
solenoidal field components), but also handles the variation of beam energy along the reference orbit.

For the cases of linear transverse coupled systems,  we consider a transverse Hamiltonnian in its most general from
\begin{equation}
H = \frac{1}{2} \left( {\bf x}^T, {\bf p}^T \right) A_c(s)
\left(
  \begin{array}{c}
     {\bf x}  \\
     {\bf p}  \\
  \end{array}
\right), ~~~
A_c(s) = \left(
        \begin{array}{cc}
         \kappa & R \\
          R^T   & m^{-1} \\
        \end{array}
      \right).
      \label{Hamiltonian}
\end{equation}
Here,
${\bf x} = (x, y)^T$ is the transverse coordinate,
${\bf p} = (p_x, p_y)^T$ is the normalized canonical momentum,
and $\kappa$ and $m^{-1}$ are $2\times2$ symmetric matrices.
The quadrupole, skew-quadrupole, and solenoidal field components are included in the focusing matrix $\kappa$,
and the relativistic mass increase along the design orbit is reflected in the mass matrix $m^{-1}$.
The arbitrary $2\times2$ matrix $R$, which is not symmetric in general, contains the solenoidal field components.
The canonical momenta are normalized by a reference momentum $p_0$, which is a constant.
The Hamiltonian equations of motion yield
\begin{eqnarray}
{\bf x'} &=& m^{-1}{\bf p} + R^T {\bf x},
\label{xp1} \\
{\bf p}' &=& - \kappa {\bf x} - R {\bf p} .
\label{xp2}
\end{eqnarray}

In Refs. \cite{GCS_PRL, GCS.PRSTAB},
the second-order matrix differential equation for the $2 \times 2$ matrix  $w$ was originally derived as
\begin{equation}
\frac{d}{d s} \left( \frac{dw}{ds} m - w Rm  \right) + \frac{dw}{ds} m R^T + w ( \kappa -R m R^T) = \left( w^T w m w^T \right)^{-1},
\end{equation}
which is the generalization of Eq. (\ref{1D_envelope}).
Here, we express it in terms of two first-order equations as
\begin{eqnarray}
W' &=& m^{-1} V + R^T W, \label{W'}\\
V' &=& - \kappa  W - R V + \left( W^T m W W^T \right)^{-1}. \label{V'}
\end{eqnarray}
where the $2\times2$ matrices $W$ and $V$ are defined by
$W = w^T$ and  $V = m \left( W'  - R^T W \right)$, respectively.
The variable $V$ can be considered to be the matrix associated with the envelope momentum \cite{Accelerator_physics3}.
We note that Eqs. (\ref{W'}) and (\ref{V'}) have  Hamiltonian structure similar to the single-particle equations of motion
(\ref{xp1}) and (\ref{xp2})
except for  the term $\left( W^T m W W^T \right)^{-1}$ .
Similar to Eq. (\ref{M(s)_1D}) in the original CS theory,
the solution of Eqs. (\ref{xp1}) and (\ref{xp2}) is expressed in terms of a symplectic linear map as \cite{GCS_PRL,GCS.PRSTAB}
\begin{equation}
\left(
\begin{array}{c}
     {\bf x}  \\
     {\bf p}  \\
  \end{array}
\right)
= \left(
      \begin{array}{cc}
        W & 0 \\
        V & W^{-T} \\
      \end{array}
    \right)
    P^{-1}
    \left(
      \begin{array}{cc}
        W^{-1} & 0 \\
        - V^T & W^T \\
      \end{array}
    \right)_0
\left(
    \begin{array}{c}
     {\bf x}  \\
     {\bf p}  \\
  \end{array}
\right)_0,
    \label{M_compact}
\end{equation}
and the 4D symplectic rotation matrix $P$ is determined by
\begin{equation}
P '
    = P \left(
          \begin{array}{cc}
            0            & - \left( W^T m W \right)^{-1} \\
            \left( W^T m W \right)^{-1} & 0 \\
          \end{array}
        \right),
\end{equation}
where $\left( W^T m W \right)^{-1}$ represents the phase advance rate.

To obtain insight on the elegant connection between the original and generalized CS theories,
we now seek to find an envelope Hamiltonian $H_{env}$, which generates  Eqs. (\ref{W'}) and (\ref{V'}) according to
\begin{eqnarray}
W' &=&    \frac{\partial H_{env}}{ \partial V},  \\
V' &=&   -  \frac{\partial H_{env}}{ \partial W}.
\end{eqnarray}
The definition of the derivative with respect to a matrix is given in Appendix.
We make a guess that the envelope Hamiltonian is composed of two contributions:
one corresponding to the quadratic terms ($H_Q$), and the other corresponding to  the higher-order nonlinear terms ($H_N$).
Motivated by the several matrix identities associated with the trace operation,
we try the following form for $H_Q$:
\begin{equation}
H_{Q} = \frac{1}{2} {\rm Tr} \left[ \left(W^T, V^T \right)
A_c(s)
\left(
        \begin{array}{c}
         W \\
         V \\
        \end{array}
\right)
\right].
\end{equation}
Explicitly, one obtains
\begin{equation}
H_{Q}
=
\frac{1}{2} {\rm Tr}
\left[
        W^T \kappa W +  W^T R V +
          V^T R^T W  +  V^T m^{-1} V
\right].
\end{equation}
Therefore, we reproduce the linear terms in Eqs. (\ref{W'}) and (\ref{V'}) as
\begin{eqnarray}
\frac{\partial H_{Q}}{ \partial V}
&=&
\frac{1}{2}
\left[
R^T W + R^T W + m^{-1} V + m^{-T} V
\right] \nonumber \\
&=&  m^{-1} V + R^T W,
\end{eqnarray}
and
\begin{eqnarray}
- \frac{\partial H_{Q}}{ \partial W}
&=&
- \frac{1}{2}
\left[
\kappa W + \kappa^T W + R V + R V 
\right] \nonumber \\
&=& - \kappa W  - R V.
\end{eqnarray}
Here, several of the matrix identities in Appendix have been applied.

Next, we seek to find the nonlinear part of the Hamiltonian $H_N$.
We note the following remarkable matrix identity \cite{cookbook}.
Assuming $C$ is symmetric, it then follows that
\begin{equation}
\frac{\partial}{\partial X} {\rm Tr} \left[ \left( X^T C X \right)^{-1} A \right] \\
= - \left[  CX (X^T C X)^{-1} \right] (A+A^T) (X^T C X)^{-1}.
\label{nonlinear}
\end{equation}
We set $A = I$ (the identity matrix), and $C = m$ (the mass matrix which is symmetric by definition) in Eq. (\ref{nonlinear}).
It then follows that
\begin{eqnarray}
\frac{\partial}{\partial X} {\rm Tr} \left[ \left( X^T m X \right)^{-1}  \right]
&=& - mX  (mX)^{-1} X^{-T} ( 2 I ) (X^T m X)^{-1} \nonumber \\
&=& - 2 X^{-T}(X^T m X)^{-1} \nonumber \\
&=& - 2 \left( X^T m X X^T \right)^{-1}.
\end{eqnarray}
If we set $X = W$ and rearrange the terms, we finally obtain,
\begin{eqnarray}
\left( W^T m W W^T \right)^{-1}
&=&      -  \frac{\partial}{\partial W} \frac{1}{2} {\rm Tr} \left[ \left( W^T m W \right)^{-1}  \right] \\
&\equiv& -  \frac{\partial H_N}{\partial W}.
\end{eqnarray}
Here, we have defined
\begin{equation}
H_N = \frac{1}{2} {\rm Tr} \left[ \left( W^T m W \right)^{-1}  \right],
\end{equation}
which yields the nonlinear term in Eq. (\ref{V'}).

Finally, we obtain the envelope Hamiltonian as
\begin{eqnarray}
H_{env}
&=& H_Q + H_N  \nonumber \\
&=&
\frac{1}{2} {\rm Tr}
\left[
          V^T m^{-1} V + W^T R V +
          V^T R^T W    +  W^T \kappa W  +
          \left( W^T m W \right)^{-1}
\right] \nonumber  \\
&=& \frac{1}{2} {\rm Tr}
\left[ \left(W^T, V^T \right)
\left(
        \begin{array}{cc}
         \kappa & R \\
          R^T   & m^{-1} \\
        \end{array}
      \right)
\left(
        \begin{array}{c}
         W \\
         V \\
        \end{array}
\right)
\right]
+
\frac{1}{2} {\rm Tr} \left[ \left( W^T m W \right)^{-1}  \right].
\label{Henv2}
\end{eqnarray}
Furthermore, we introduce  the effective envelope potential $V_{env}$ defined as
\begin{equation}
V_{env} = \frac{1}{2} {\rm Tr}
\left[
W^T \kappa W  + \left( W^T m W \right)^{-1}
\right]
+{\rm Tr}
\left[ W^T R V \right]
,
\end{equation}
which is momentum-dependent.
We emphasize the remarkable similarities between Eqs. (\ref{Henv1}) and (\ref{Henv2}).
Indeed, Eq. (\ref{Henv2}) includes Eq. (\ref{Henv1}) as a special case.
Although we have taken $W$ and $V$ to be $2 \times 2$ matrices for most of
the derivations, the envelope Hamiltonian (\ref{Henv2}) is more general and can be readily applied to envelope equations of higher dimensions.


In summary, making use of the recently developed generalized CS theory \cite{GCS_PRL, GCS.PRSTAB},
we have formulated the envelope Hamiltonian for charged particle beam dynamics in general linear coupled lattices.
The envelope Hamiltonian reveals elegant matrix structures, and
retains all the features of the original CS theory with remarkable similarity.
We strongly expect that the discovery of the envelope Hamiltonian will provide
deeper insight into the general coupled beam dynamics,
for which no single approach has yet become standard in the accelerator physics community.

\appendix*

\section{Matrix Derivatives}
The derivative of a scalar function $f$ with respect to a matrix $X$ is defined as a matrix with the same
shape, of which elements are the partial derivatives of $f$ with respect to the
elements of $X$ \cite{MatrixAlgebra}.
Since the trace of an arbitrary square matrix $F$, ${\rm Tr}(F) = \sum_i F_{ii}$, is a scalar function,
$\partial {\rm Tr}(F) / \partial X$ is properly defined.
By making use of the following identity
\begin{equation}
\frac{ \partial X_{kl} }{ \partial X_{ij} } = \delta_{ik} \delta_{lj},
\end{equation}
and assuming that $A$ and $B$ are constant matrices, one can obtain after some straightforward algebra \cite{cookbook}
that
\begin{equation}
\frac{\partial}{\partial X} {\rm Tr} \left(  A X \right)
= A^T,
\end{equation}
\begin{equation}
\frac{\partial}{\partial X} {\rm Tr} \left( X^T A \right)
= A,
\end{equation}
\begin{equation}
\frac{\partial}{\partial X} {\rm Tr} \left( X^T B X \right)
= B X + B^T X.
\end{equation}
These matrix identities play a key role in the derivation of the envelope Hamiltonian.
We also note that the trace operation has the following useful properties;
$
{\rm Tr} \left( A B  \right) = {\rm Tr} \left( B A \right),
{\rm Tr} \left( A \right) = {\rm Tr} \left( A^T \right),
{\rm Tr} \left( A + B  \right) = {\rm Tr} \left( A \right) + {\rm Tr} \left(  B  \right)$, and
${\rm Tr} \left( a A \right) = a {\rm Tr} \left( A \right)$,
where $a$ is an arbitrary scalar.

\section*{ACKNOWLEDGMENTS}
This work was supported by the National Research Foundation of Korea (NRF-2015R1D1A1A01061074) grant funded by the Korean government
(MSIP: Ministry of Science, ICT and Future Planning).
This work was also supported by the U.S. Department of Energy Grant No. DE-AC02-09CH11466.

\end{document}